\documentclass[12pt,thmsa]{article}
\usepackage{amsfonts}

\usepackage{sw20aip}

\input{tcilatex}
\begin{document}

\author{Ron S. Rubin \\
Mathematics Department\\
Massachusetts Institute of Technology\\
rubin@math.mit.edu \and Andrzej Lesniewski \\
Paribas Capital Markets\\
New York, NY 10019\\
Andrew\_Lesniewski@paribas.com}
\title{Quantum Mechanics on a Torus\thanks{%
26 pages, 2 Figures}}
\date{June 8, 1998}
\maketitle

\begin{abstract}
We present here a canonical description for quantizing classical maps on a
torus. We prove theorems analagous to classical theorems on mixing and
ergodicity in terms of a \textit{quantum Koopman space }$L^{2}\left( \frak{A}%
_{\hbar },\tau _{\hbar }\right) $ obtained as the completion of the algebra
of observables $\frak{A}_{\hbar }$ in the norm induced by the following
inner product $\left( A,B\right) =\tau _{\hbar }\left( A^{\dagger }B\right) $%
, where $\tau _{\hbar }$ is a linear functional on the algebra analogous to
the classical ``integral over phase space.'' We also derive explicit
formulas connecting this formulation to the $\theta $-torus decomposition of
Bargmann space introduced in ref. \cite{KLMR}.
\end{abstract}

\section{\protect\smallskip Introduction}

Quantum mechanics on a toroidal phase space has reared its head in many
varied fields: from string theory (ref. \cite{Fa}) to quantum computation
(ref. \cite{Sc}), the quantized torus seems to present a simple and workable
framework for understanding the quantum world when phase space is compact.
In fact, two seemingly different approaches to the ``quantum torus'' are
widely used in the literature.

In physics papers, an ad hoc ``quantization'' is used in which a
prescription for a unitary $N\times N$ matrix (with $N=$ inverse Planck's
constant) is given which yields the classical dynamics in the limit as $%
N\rightarrow \infty $ $\left( h\rightarrow 0\right) \,$(see ref. \cite{BV1}%
). A more common approach in mathematical physics, and one which we adopt
here as our starting point, is to quantize the classical algebra generated
by $\exp \left( 2\pi ix\right) $ and $\exp \left( 2\pi ip\right) $, the
generators of the periodic functions on the plane, and hence functions on
the torus. In refs. \cite{KLMR} and \cite{E}, a connection between these two
approaches was given explicitly.

In this latter, algebraic approach, classical observables (functions) over a
toroidal phase space are replaced by non-commuting operators in a canonical
fashion: 
\begin{eqnarray*}
\exp \left( 2\pi ix\right) &\rightarrow &U \\
\exp \left( 2\pi ip\right) &\rightarrow &V
\end{eqnarray*}
where $U$ and $V$ satisfy the canonical commutation relation 
\[
UV=e^{4\pi ^{2}i\hbar }VU. 
\]
The Hilbert space on which these operators act is taken to be the usual $%
L^{2}\left( \Bbb{R},dx\right) $ or, equivalently, Bargmann space $\mathcal{H}%
^{2}\left( \mathbb{C},d\mu _{\hbar }\right) \,$(see below). Corresponding to
any classical function over the torus $f$ we specify a quantum operator $%
Q_{\hbar }\left( f\right) .$ That is, $Q_{\hbar }$ also gives an \textit{%
ordering }prescription for the quantum operator.

We say $\{U,$ $V,U^{-1},V^{-1}\}$ (operators on $\mathcal{H}^{2}\left( %
\mathbb{C},d\mu _{\hbar }\right) $) generate the quantum algebra of
observables $\frak{A}_{\hbar }$. The only appropriate ``quantum mechanical
questions'' must be operators which depends on $U$ and $V$. For example, it
is appropriate to ask: \textit{What is }$\left( U+U^{\dagger }\right) /2$%
\textit{\ (an observable)?} Whereas, the question: \textit{What is }$\exp
\left( -\widehat{x}^{2}\right) $\textit{?} is not allowed. Different
closures on series generated by these operators and their inverses will
yield different algebras: the smallest are the $\mathbb{C}^{*}$-algebras
which are norm closed, while von Neumann algebras are only weakly closed.
The quantization of classical maps can proceed in a such a fashion for all
values of Planck's constant. In this framework, a natural quantum version of
chaos can be defined. The procedure is, simply: formulate the classical
definition algebraically, and substitute the quantum noncommutative algebra
for the classical algebra.

The dynamical component of the quantization involves finding the (highly
non-unique) unitary quantum propagator $F$ such that 
\begin{equation}
\lim_{\hbar \rightarrow 0}F^{\dagger }Q_{\hbar }\left( f\right) F=Q_{\hbar
}\left( f\circ T\right)  \label{classical limit}
\end{equation}
in a suitable topology. Here $T$ represents the classical map on the torus.
What this means is we require the propagator to return the classical
dynamics in the limit $\hbar \rightarrow 0$. As is common in the literature,
we write the quantum evolution as an action of the group $\mathbb{Z}$ on $%
\frak{A}_{\hbar }$. That is, for any $A\in \frak{A}_{\hbar }$, 
\[
\alpha _{n}\left( A\right) =F^{-n}AF^{n}\text{.} 
\]

Finally, for discussions of quantum ergodicity, we define a ``quantum
ensemble average'' $\tau _{\hbar }$ as a linear functional on $\frak{A}%
_{\hbar }.$ The defining property of this functional is the same as the
classical ``phase space average'': 
\begin{eqnarray*}
\text{classical} &:&\text{{}}\int_{T^{2}}e^{2\pi i\left( mx+np\right)
}dxdp=\delta _{m0}\delta _{n0}, \\
\text{quantum} &:&\text{ }\tau _{\hbar }\left( U^{m}V^{n}\right) =\delta
_{m0}\delta _{n0}.
\end{eqnarray*}

We summarize this discussion with the following general definition:

\begin{definition}
\label{quantum map}A \textit{quantum map} is a quadruple $\left( \frak{A}%
_{\hbar },\mathbb{Z},\alpha _{n},\tau _{\hbar }\right) $ (or $\left( \frak{A}%
_{\hbar },\mathbb{Z},F,\tau _{\hbar }\right) $) such that

(i) $\frak{A}_{\hbar }$ is an algebra of bounded operators.

(ii) $\alpha _{n}:$ $\mathbb{Z\rightarrow }Aut\left( \frak{A}_{\hbar
}\right) $ is an action of $\mathbb{Z}$ on $\frak{A}_{\hbar }$.

(iii) $\tau _{\hbar }$ is a $\mathbb{Z}$-invariant faithful tracial state on 
$\frak{A}_{\hbar }$: 
\begin{equation}
\tau _{\hbar }\left( \alpha _{n}\left( A\right) \right) =\tau _{\hbar
}\left( A\right) \text{.}  \label{Tau invariance}
\end{equation}
\end{definition}

Recall that a \textit{classical map} may be defined as the triplet $\left(
M,\mu ,T_{n}\right) $, with $M$ a smooth manifold, $\mu $ a positive measure
on $M$, and $T_{n}$ a one-parameter group of measure-preserving
diffeomorphisms, with $n\in \mathbb{Z}$. This leads to a natural definition
of what it means to a have a ``quantization of a classical dynamics'':

\begin{definition}
The quantum map $\left( \frak{A}_{\hbar },\mathbb{Z},\alpha _{n},\tau
_{\hbar }\right) $is furthermore the quantization of a classical map $\left(
M,\mu ,T_{n}\right) $ if, for any $f\in $ $\frak{A}_{0}$ (the classical
algebra of functions on $M$), 
\begin{eqnarray*}
\frak{A}_{\hbar } &=&Q_{\hbar }\left( \frak{A}_{0}\right) \text{,} \\
\alpha _{n}\left( Q_{\hbar }\left( f\right) \right) &\rightarrow &Q_{\hbar
}\left( f\circ T_{n}\right) \quad \text{as\quad }\hbar \rightarrow 0, \\
\tau _{\hbar }\left( Q_{\hbar }\left( f\right) \right) &\rightarrow
&\int_{M}fd\mu \quad \text{as\quad }\hbar \rightarrow 0\text{.}
\end{eqnarray*}
\end{definition}

An important point is that Definition \ref{quantum map} for a quantum map
can be ``weakened'' by relaxing conditions (ii) and (iii) to hold in the
limit $\hbar \rightarrow 0$.

\section{Classical Maps and Classical Ergodicity}

Much work has recently been done in the field of classical chaos. In
particular, a hierarchy of just how ``chaotic'' a dynamics can be has been
established: 
\[
\text{Anosov Dynamics}\Rightarrow \text{Mixing}\Rightarrow \text{Ergodic}. 
\]

A system is defined to be Anosov if the phase space is compact and the flow
is globally hyperbolic. A canonical example of such a system is the dynamics
of a particle constrained to a compact surface of negative curvature with
genus $g\geq 2$ (see \cite{BV1}). In such a system two particles which are
nearby almost always fall away from each other exponentially fast. This,
plus the fact the dynamics is constrained to a compact phase space induces
the strong form of chaos known as the Anosov property. It is remarkable just
how ``random'' this motion can get. Anosov systems satisfy the Bernoulli
property, in which two points initially close to each other will eventually
be as different as two different infinite series of coin tosses. (See ref. 
\cite{M} for a full discussion, and ref. \cite{OW} for a proof.) A beautiful
example of a map which satisfies this Bernoulli property is the baker's map,
whose quantization is studied in refs. \cite{BV1} and \cite{RS}.

A weaker notion of chaos is known as the mixing property. (It was shown by
Sinai in ref. \cite{S} that all Anosov systems are mixing.) A dynamical
system is mixing if all areas of initial particles eventually spread
uniformly throughout phase space. More precisely, a map $\left( M,\mu
,T_{n}\right) $ is \textit{mixing} if for every $f,g\in L^{2}\left( M,\mu
\right) $ 
\begin{equation}
\lim_{n\rightarrow \infty }\int_{\mu }\overline{f\left( T^{n}\left( x\right)
\right) }g\left( x\right) d\mu \left( x\right) =\int_{\mu }f\left( x\right)
d\mu \left( x\right) \int_{\mu }g\left( x\right) d\mu \left( x\right) .
\label{mixing}
\end{equation}

A basic result of ergodic theory is that mixing implies ergodicity.

\begin{definition}
\label{ergodicity}A system is ergodic if any invariant measurable function $%
f\in L^{2}\left( M,\mu \right) $ is constant almost everywhere.
\end{definition}

A more common (and physically clear), but equivalent definition of ergodic
is ``time average $=$ phase space average''. More precisely, a map $\left(
M,\mu ,T_{n}\right) $ is ergodic if for any $f\in L^{2}\left( M,\mu \right) $%
, 
\[
s-\lim_{M\rightarrow \infty }\frac{1}{M}\sum_{m=0}^{M-1}f\circ
T^{M}=\int_{\mu }f\left( x\right) d\mu \left( x\right) . 
\]

Ergodicity is weaker than mixing in that a measurable set can actually
remain intact in time while individual points still visit the entire phase
space. That is, a ``packet'' can survive. A good example of a system which
is ergodic but not mixing is the Kronecker map (see ref. \cite{KLMR}).

\section{Quantum Ergodicity}

We can associate with the quadruple $\left( \frak{A}_{\hbar },\mathbb{Z}%
,\alpha _{n},\tau _{\hbar }\right) $ a quantum Koopman space as a Hilbert
space of operators derived from $\frak{A}_{\hbar }$.

\begin{definition}
The \textit{quantum Koopman space }is the Hilbert space $L^{2}\left( \frak{A}%
_{\hbar },\tau _{\hbar }\right) $ obtained as the completion of $\frak{A}%
_{\hbar }$ in the norm induced by the following inner product 
\[
\left( A,B\right) =\tau _{\hbar }\left( A^{\dagger }B\right) \text{,} 
\]
for $A,B$ $\in $ $\frak{A}_{\hbar }.$
\end{definition}

What we shall now demonstrate that this ``quantum Hilbert space'' $%
L^{2}\left( \frak{A}_{\hbar },\tau _{\hbar }\right) $ has associated with it
many of the theorems in ergodicity that are classically associated with $%
L^{2}\left( M,\mu \right) $. For clarity, we give here explicit proofs for
the case $G=\mathbb{Z}$ of quantum maps, although the results can easily be
generalized to the case of any amenable group.

First, we observe that the quantum evolution is unitary on $L^{2}\left( 
\frak{A}_{\hbar },\tau _{\hbar }\right) $. Specifically, the automorphism $%
\alpha $ which implements the quantum dynamics on $\frak{A}_{\hbar }$ is
unitary on $L^{2}\left( \frak{A}_{\hbar },\tau _{\hbar }\right) $. This
follows almost immediately from eqn. \ref{Tau invariance}. We see 
\begin{eqnarray*}
\left( A,\alpha \left( B\right) \right) &=&\tau _{\hbar }\left( A^{\dagger
}\alpha \left( B\right) \right) \\
&=&\tau _{\hbar }\left( \alpha _{-1}\left( A^{\dagger }\alpha \left(
B\right) \right) \right) \\
&=&\tau _{\hbar }\left( \alpha _{-1}\left( A^{\dagger }\right) B\right)
\end{eqnarray*}
Thus $\alpha _{-1}=\alpha ^{\dagger }$ .

We can think of $\alpha $ as the ``quantum Liouville operator'' analogous to
the unitary Liouville operator of classical dynamics.

We now define ``quantum ergodicity'' in analogy with the classical
definition \ref{ergodicity}.

\begin{definition}
\label{quantergodic}A quantum map is called \textit{ergodic} if the only $%
A\in L^{2}\left( \frak{A}_{\hbar },\tau _{\hbar }\right) $ invariant under $%
\mathbb{Z}$ are scalar multiples of the identity $I$.\smallskip
\end{definition}

In what sense is this a statement of ``time average $=$ phase space
average''? We first demonstrate the existence of the ``quantum time
average,'' analogous to the von Neuman theorem of classical dynamical
systems (see ref. \cite{H}).

\begin{theorem}
For any $A\in L^{2}\left( \frak{A}_{\hbar },\tau _{\hbar }\right) $, the
strong limit 
\[
s-\lim_{n\rightarrow \infty }\frac{1}{N}\sum_{n=0}^{N-1}\alpha _{n}\left(
A\right) =A_{*} 
\]
exists.
\end{theorem}

Proof. We divide the proof into three steps.

Step1. Let $A$ $\in L^{2}\left( \frak{A}_{\hbar },\tau _{\hbar }\right) $
have the form 
\[
A=B-\alpha _{m}\left( B\right) 
\]
where $B\in \frak{A}_{\hbar }$. Observe that 
\begin{eqnarray*}
\frac{1}{N}\sum_{n=0}^{N-1}\alpha _{n}\left( A\right) &=&\frac{1}{N}%
\sum_{n=0}^{N-1}\alpha _{n}\left( B-\alpha _{m}\left( B\right) \right) \\
&=&\frac{1}{N}\sum_{n=0}^{N-1}\alpha _{n}\left( B\right) -\alpha
_{m+n}\left( B\right) \\
&=&\frac{1}{N}\left( \sum_{n=0}^{N-1}\alpha _{n}\left( B\right)
-\sum_{n=m}^{N+m-1}\alpha _{n}\left( B\right) \right) . \\
&=&\frac{1}{N}\left( \sum_{n=-\infty }^{\infty }\chi _{[0,N-1]}\left(
n\right) \alpha _{n}\left( B\right) -\sum_{n=-\infty }^{\infty }\chi
_{[m,m+N-1]}\left( n\right) \alpha _{n}\left( B\right) \right) \\
&=&\frac{1}{N}\left( \sum_{n=-\infty }^{\infty }\left( \chi _{[0,N-1]}\left(
n\right) -\chi _{[m,m+N-1]}\left( n\right) \right) \alpha _{n}\left(
B\right) \right)
\end{eqnarray*}
where $\chi _{[a,b]}$ is the characteristic function 
\[
\chi _{[a,b]}\left( n\right) =\left\{ 
\begin{array}{ll}
1, & a\leq n\leq b \\ 
0, & \text{otherwise}.
\end{array}
\right. 
\]
Then 
\begin{eqnarray*}
\left\| \frac{1}{N}\sum_{n=0}^{N-1}\alpha _{n}\left( A\right) \right\|
_{L^{2}} &\leq &\frac{1}{N}\sum_{n=-\infty }^{\infty }\left| \left( \chi
_{[0,N-1]}\left( n\right) -\chi _{[m,m+N-1]}\left( n\right) \right) \right|
\left\| \alpha _{n}\left( B\right) \right\| _{L^{2}} \\
&=&\left\| B\right\| _{L^{2}}\frac{1}{N}\sum_{n=-\infty }^{\infty }\left|
\left( \chi _{[0,N-1]}\left( n\right) -\chi _{[m,m+N-1]}\left( n\right)
\right) \right|
\end{eqnarray*}
Now choose $N>m$. Then 
\begin{eqnarray*}
\sum_{n=-\infty }^{\infty }\left| \left( \chi _{[0,N-1]}\left( n\right)
-\chi _{[m,m+N-1]}\left( n\right) \right) \right| &=&\sum_{n=-\infty
}^{m-1}\left| \left( \chi _{[0,N-1]}\left( n\right) -\chi _{[m,m+N-1]}\left(
n\right) \right) \right| \\
&&+\sum_{n=m}^{N-1}\left| \left( \chi _{[0,N-1]}\left( n\right) -\chi
_{[m,m+N-1]}\left( n\right) \right) \right| \\
&&+\sum_{n=N}^{N+m-1}\left| \left( \chi _{[0,N-1]}\left( n\right) -\chi
_{[m,m+N-1]}\left( n\right) \right) \right| \\
&&+\sum_{n=N+m-1}^{\infty }\left| \left( \chi _{[0,N-1]}\left( n\right)
-\chi _{[m,m+N-1]}\left( n\right) \right) \right| \\
&=&2m.
\end{eqnarray*}
so that 
\[
\left\| \frac{1}{N}\sum_{n=0}^{N-1}\alpha _{n}\left( A\right) \right\|
_{L^{2}}\leq \left\| B\right\| _{L^{2}}\frac{2m}{N}\rightarrow 0\text{\quad
as\quad }N\rightarrow \infty \text{.} 
\]

Step 2. We denote the closed subspace in $L^{2}\left( \frak{A}_{\hbar },\tau
_{\hbar }\right) $ generated by operators of the form $B-\alpha _{m}\left(
B\right) $ as $\mathcal{H}$. Then for any $A\in \mathcal{H}$, and any $%
\epsilon $, there exists an $A_{0}=$ $B-\alpha _{m}\left( B\right) $ such
that 
\[
\left\| A-A_{0}\right\| _{L^{2}}\leq \epsilon /2. 
\]
Thus 
\begin{eqnarray*}
\left\| \frac{1}{N}\sum_{n=0}^{N-1}\alpha _{n}\left( A\right) \right\|
_{L^{2}} &\leq &\left\| \frac{1}{N}\sum_{n=0}^{N-1}\alpha _{n}\left(
A_{0}\right) \right\| _{L^{2}}+\left\| \frac{1}{N}\sum_{n=0}^{N-1}\alpha
_{n}\left( A-A_{0}\right) \right\| _{L^{2}} \\
&\leq &\left\| \frac{1}{N}\sum_{n=0}^{N-1}\alpha _{n}\left( A_{0}\right)
\right\| _{L^{2}}+\left\| A-A_{0}\right\| _{L^{2}}
\end{eqnarray*}
By the results of step 1, we can choose an $N$ such that 
\[
\left\| \frac{1}{N}\sum_{n=0}^{N-1}\alpha _{n}\left( A_{0}\right) \right\|
_{L^{2}}\leq \epsilon /2 
\]
so that for all $A\in \mathcal{H}$, 
\[
\left\| \frac{1}{N}\sum_{n=0}^{N-1}\alpha _{n}\left( A\right) \right\|
_{L^{2}}\rightarrow 0\text{\quad as\quad }N\rightarrow \infty \text{.} 
\]

Step 3. By the Riesz lemma (see \cite{RS}), any $A\in L^{2}\left( \frak{A}%
_{\hbar },\tau _{\hbar }\right) $ can be written $A=A^{\parallel }+A^{\perp
} $, where $A^{\parallel }\in \mathcal{H}$ and $A^{\perp }$ is in the
orthogonal complement $\mathcal{H}^{\perp }$. Observe that for any $B\in
L^{2}\left( \frak{A}_{\hbar },\tau _{\hbar }\right) $, 
\begin{eqnarray*}
0 &=&\left( A^{\perp },B-\alpha _{m}\left( B\right) \right) =\left( \alpha
_{-m}\left( A^{\perp }\right) ,\alpha _{-m}\left( B\right) -B\right) \\
&=&\left( \alpha _{-m}\left( A^{\perp }\right) ,\alpha _{-m}\left( B\right)
\right) -\left( \alpha _{-m}\left( A^{\perp }\right) ,B\right) \\
&=&\left( A^{\perp }-\alpha _{-m}\left( A^{\perp }\right) ,B\right) .
\end{eqnarray*}
Thus, for any $m$%
\[
\left( A^{\perp },B\right) =\left( \alpha _{m}\left( A^{\perp }\right)
,B\right) . 
\]
from which it follows that $A^{\perp }=\alpha _{m}\left( A^{\perp }\right) $%
. Thus 
\[
\frac{1}{N}\sum_{n=0}^{N-1}\alpha _{n}\left( A^{\perp }\right) =A^{\perp } 
\]
and the theorem is proved. 
\endproof%

Another important result is that the time average is in fact \textit{%
invariant} under the dynamics.

\begin{lemma}
\label{time invariance}For any $n\in \mathbb{Z}$, $\alpha _{n}\left(
A_{*}\right) =A_{*}$.
\end{lemma}

Proof. Observe that 
\begin{eqnarray*}
\alpha _{n}\left( A_{*}\right) &=&s-\lim_{M\rightarrow \infty }\frac{1}{M}%
\sum_{m=0}^{M-1}\alpha _{m+n}\left( A\right) \\
&=&s-\lim_{M\rightarrow \infty }\frac{1}{M}\sum_{m=n}^{M+n-1}\alpha
_{m}\left( A\right) \\
&=&\left( s-\lim_{M\rightarrow \infty }\frac{1}{M}\sum_{m=0}^{M-1}\alpha
_{m}\left( A\right) \right) -\left( s-\lim_{M\rightarrow \infty }\frac{1}{M}%
\sum_{m=0}^{n-1}\alpha _{m}\left( A\right) \right) \\
&&+\left( s-\lim_{M\rightarrow \infty }\frac{1}{M}\sum_{m=M}^{M+n-1}\alpha
_{m}\left( A\right) \right) \\
&=&A_{*}-\left( s-\lim_{M\rightarrow \infty }\frac{1}{M}\sum_{m=0}^{n-1}%
\alpha _{m}\left( A\right) \right) +\left( s-\lim_{M\rightarrow \infty }%
\frac{1}{M}\sum_{m=M}^{M+n-1}\alpha _{m}\left( A\right) \right) .
\end{eqnarray*}
Now consider the middle term. We have 
\begin{eqnarray*}
\left\| s-\lim_{M\rightarrow \infty }\frac{1}{M}\sum_{m=0}^{n-1}\alpha
_{m}\left( A\right) \right\| _{L^{2}} &\leq &\lim_{M\rightarrow \infty }%
\frac{1}{M}\sum_{m=0}^{n-1}\left\| \alpha _{m}\left( A\right) \right\|
_{L^{2}} \\
&=&\lim_{M\rightarrow \infty }\frac{n}{M}\left\| A\right\| _{L^{2}}=0\text{.}
\end{eqnarray*}
by the unitarity of $\alpha $. The same argument holds for the last term,
and the lemma is thus proved.%
\endproof%

We are now in a position to derive the more conceptual statement of quantum
ergodicity, whereby the time average equals the phase space average.

\begin{theorem}
A quantum map $\left( \frak{A}_{\hbar },\alpha ,\mathbb{Z},\tau _{\hbar
}\right) $ is quantum ergodic if and only if 
\begin{equation}
s-\lim_{n\rightarrow \infty }\frac{1}{N}\sum_{n=0}^{N-1}\alpha _{n}\left(
A\right) =\tau _{\hbar }\left( A\right) I  \label{quantum ergodicity}
\end{equation}
for all $A\in \frak{A}_{\hbar }$.
\end{theorem}

Proof. For an ergodic quantum map, we see from Definition \ref{quantergodic}
and Lemma \ref{time invariance} that 
\[
s-\lim_{n\rightarrow \infty }\frac{1}{N}\sum_{n=0}^{N-1}\alpha _{n}\left(
A\right) =c_{A}I\text{.} 
\]
We next apply $\tau _{\hbar }$ to both sides to get 
\[
s-\lim_{n\rightarrow \infty }\frac{1}{N}\sum_{n=0}^{N-1}\tau _{\hbar }\left(
\alpha _{n}\left( A\right) \right) =c_{A}\text{.} 
\]
By the invariance of $\tau _{\hbar }$, we see immediately that $c_{A}=\tau
_{\hbar }\left( A\right) $. Conversely, if \ref{quantum ergodicity} holds,
then suppose $\alpha _{n}\left( A\right) =A$ for some $A\in \frak{A}_{\hbar
} $. Then 
\[
s-\lim_{n\rightarrow \infty }\frac{1}{N}\sum_{n=0}^{N-1}\alpha _{n}\left(
A\right) =A=\tau _{\hbar }\left( A\right) I, 
\]
i.e. $A$ is a multiple of the identity operator. 
\endproof%

We can also formulate an analogous definition for \textit{quantum mixing.}

\begin{definition}
A quantum map $\left( \frak{A}_{\hbar }\text{, }\alpha \text{, }\mathbb{Z}%
\text{, }\tau _{\hbar }\right) $ is called mixing if for all $A,B\in \frak{A}%
_{\hbar }$%
\begin{equation}
\lim_{n\rightarrow \infty }\tau _{\hbar }\left( \alpha _{n}\left( A\right)
B\right) =\tau _{\hbar }\left( A\right) \tau _{\hbar }\left( B\right) \text{.%
}  \label{quantum mix}
\end{equation}
\end{definition}

Comparison of this with \ref{mixing} demonstrates that \ref{quantum mix} is
indeed a quantum mechanical version of mixing. As in the classical case,
quantum mixing is a stronger statement than quantum ergodicity.

\begin{theorem}
\label{mixing implies ergodic} For a quantum map $\left( \frak{A}_{\hbar }%
\text{, }\alpha \text{, }\mathbb{Z}\text{, }\tau _{\hbar }\right) $, quantum
mixing implies quantum ergodicity.
\end{theorem}

Proof. Consider the case where $\tau _{\hbar }\left( A\right) =0$. Observe
that 
\begin{eqnarray*}
\left\| \frac{1}{N}\sum_{n=0}^{N-1}\alpha _{n}\left( A\right) \right\| ^{2}
&=&\tau _{\hbar }\left( \left( \frac{1}{N}\sum_{n=0}^{N-1}\alpha _{n}\left(
A\right) \right) ^{\dagger }\left( \frac{1}{M}\sum_{m=0}^{M-1}\alpha
_{m}\left( A\right) \right) \right) \\
&=&\frac{1}{NM}\sum_{n=0}^{N-1}\sum_{m=0}^{M-1}\tau _{\hbar }\left( \alpha
_{n}\left( A^{\dagger }\right) \alpha _{m}\left( A\right) \right) \\
&=&\frac{1}{NM}\sum_{n=0}^{P-1}\sum_{m=0}^{Q-1}\tau _{\hbar }\left( \alpha
_{n}\left( A^{\dagger }\right) \alpha _{m}\left( A\right) \right) \\
&&+\frac{1}{NM}\sum_{n=P}^{N-1}\sum_{m=Q}^{M-1}\tau _{\hbar }\left( \alpha
_{n}\left( A^{\dagger }\right) \alpha _{m}\left( A\right) \right) \\
&\leq &\frac{PQ}{NM}\left\| A\right\| ^{2}+\frac{1}{NM}\sum_{n=P}^{N-1}%
\sum_{m=Q}^{M-1}\tau _{\hbar }\left( \alpha _{n}\left( A^{\dagger }\right)
\alpha _{m}\left( A\right) \right) \\
&\leq &\frac{PQ}{NM}\left\| A\right\| ^{2}+\sup_{P\geq N,Q\geq M}\left| \tau
_{\hbar }\left( \alpha _{P}\left( A^{\dagger }\right) \alpha _{Q}\left(
A\right) \right) \right| .
\end{eqnarray*}
for some fixed $P\leq N$ and $Q\leq M$. By the definition of mixing, we can
choose a $P$ and $Q$ such that the second set of terms factorizes, i.e. is
less than $\epsilon /2$ for any $\epsilon $. For this $P$ and $Q$, we can
now find an $M\,$ and $N$ such that the first term is less than $\epsilon /2$%
. Thus, for any $\epsilon $, and sufficiently large $N$, 
\begin{equation}
\left\| \frac{1}{N}\sum_{n=0}^{N-1}\alpha _{n}\left( A\right) \right\|
^{2}\leq \epsilon \text{.}  \label{ergodic inequality}
\end{equation}
This proves that mixing implies ergodicity for the case $\tau _{\hbar
}\left( A\right) =0$. For the general case, we simply let $A_{0}=A-\tau
_{\hbar }\left( A\right) $ in \ref{ergodic inequality}. This completes the
proof of the theorem. 
\endproof%

What we see here is that many of the connections the different formulations
of ergodicity in classical mechanics remain true quantum mechanically. This
provides an impetus to believe that quantum mechanics does preserve notions
of chaotic dynamics, despite the uncertainty principle. What's more, it also
gives hope that a physically sound and well-accepted hierarchy of ``quantum
chaos'' is within reach.

\section{The Quantum Torus}

\subsection{Bargmann Space}

The classical dynamics on a toroidal phase space are fundamentally phase
space dynamics, that is, the momentum coordinate is not necessarily the
symplectic dual of the position coordinate. This leads to the use of a
Bargmann representation of Hilbert space. (See \cite{F} for a comprehensive
review of Bargmann space.) Bargmann space $\mathcal{H}^{2}\left( \mathbb{C}%
,d\mu _{\hbar }\right) $ is the Hilbert space of entire functions on the
complex plane which are square integrable with respect to the measure $d\mu
_{\hbar }\left( z\right) =\left( \pi \hbar \right) ^{-1}\exp \left( -\left|
z\right| ^{2}/\hbar \right) .$ In fact, as we shall presently review, this
is equivalent to $L^{2}\left( \mathbb{R},dx\right) $ via the Bargmann
transform. The idea behind Bargmann space is to project a general state $%
\left| \psi \right\rangle $ onto a coherent state $\left| \overline{z}%
\right\rangle $ then multiply out the non-analytic part to obtain an entire
function on the complex plane. Thus, we write 
\[
\psi \left( z\right) =e^{\left| z\right| ^{2}/2\hbar }\left\langle \overline{%
z}\right| \left. \psi \right\rangle . 
\]
In fact, there is an inner product isomorphism between $\mathcal{H}^{2}(%
\mathbb{C},d\mu _{\hbar })$ and $L^{2}(\mathbb{R},dx)$, implemented by the
Bargmann transform. For $x\in \mathbb{R}$, and $z\in \mathbb{C}$, and using
an appropriate normalization, this can be found easily 
\begin{eqnarray*}
\psi \left( x\right) &=&\left\langle x\right| \left. \psi \right\rangle =%
\frac{1}{\pi \hbar }\int_{\mathbb{C}}\left\langle x\right| \left. \overline{z%
}\right\rangle \left\langle \overline{z}\right| \left. \psi \right\rangle
d^{2}z \\
&=&\left( \pi \hbar \right) ^{-1/4}\int_{\mathbb{C}}\exp \left( \left( \sqrt{%
2}xz-x^{2}/2-z^{2}/2\right) /\hbar \right) \psi \left( z\right) d\mu _{\hbar
}\left( z\right) .
\end{eqnarray*}
Hence, 
\begin{eqnarray}
B &:&\mathcal{H}^{2}(\mathbb{C},d\mu _{\hbar })\rightarrow L^{2}(\mathbb{R}%
,dx),  \label{bargmann transformation} \\
B(z,x) &=&\frac{e^{\left( \sqrt{2}xz-x^{2}/2-z^{2}/2\right) /\hbar }}{(\pi
\hbar )^{1/4}},  \nonumber
\end{eqnarray}
and $B^{-1}(z,x)=B(\overline{z},x)$. We now review some of the important
properties of Bargmann space. First of all, the polynomials in $z$
(obviously entire) span a countably dense subset, and in fact, $%
1,z,z^{2},z^{3},...$ are orthogonal functions. We can see this easily by
using spherical coordinates with $z=re^{i\phi }$: 
\[
\left( z^{n},z^{m}\right) =\int_{\mathbb{C}}\overline{z}^{n}z^{m}d\mu
_{\hbar }\left( z\right) =2^{-n}\pi ^{1/2}\hbar ^{n-1/2}\left( 2n-1\right)
!!\delta _{nm}. 
\]
Bargmann representation also carries a unitary projective representation of
the group of translations on the plane. Let $a$ and $b$ be fixed complex
numbers and let $\psi \in \mathcal{H}^{2}(\mathbb{C},d\mu _{\hbar })$. Then 
\begin{equation}
U\left( a\right) \psi \left( z\right) =\exp \left( \frac{1}{\hbar }\left( 
\overline{a}z-\left| a\right| ^{2}/2\right) \right) \phi \left( z-a\right) ,
\label{translations}
\end{equation}
so that 
\begin{equation}
U\left( a\right) U\left( b\right) =e^{i\func{Im}\left( \overline{a}b/\hbar
\right) }U\left( a+b\right) .  \label{projective rep}
\end{equation}

Also important is the fact that Bargmann space has a reproducing kernel $%
\exp \left( \overline{w}z/\hbar \right) $ satisfying the equation 
\begin{equation}
\int_{\mathbb{C}}e^{z\overline{w}/\hbar }\psi \left( w\right) d\mu _{\hbar
}\left( w\right) =\psi \left( z\right) .  \label{reproducing kernel}
\end{equation}

Note that the fact that it is a projective representation of the group of
translations rather than a representation is a manifestation of the
uncertainty principle, where we interpret the real part of $z$ as the
canonical position coordinate and the imaginary part as the canonical
momentum. With this in mind, we next define the particular quantization
ordering we shall use throughout. We define $z=\left( x-ip\right) /\sqrt{2}$
with $x$ and $p$ $\in \mathbb{R}$. Corresponding to a classical symbol $%
\sigma $ (a function over phase space) we define the quantum observable as
the Toeplitz operator $T_{\hbar }\left( \sigma \right) $ acting on $\mathcal{%
H}^{2}(\mathbb{C},d\mu _{\hbar })$ so that 
\begin{equation}
T_{\hbar }\left( \sigma \right) \psi \left( z\right) =\int_{\mathbb{C}}e^{z%
\overline{w}/\hbar }\sigma \left( w\right) \psi \left( w\right) d\mu _{\hbar
}\left( w\right) .  \label{Toeplitz}
\end{equation}
Observe some of the remarkable properties of this quantization. First of
all, acting on a symbol which is entire over the plane, it is just
multiplication by the symbol. In particular, 
\[
\widehat{z}\psi \left( z\right) \equiv T_{\hbar }\left( z\right) \psi \left(
z\right) =z\psi \left( z\right) . 
\]
The quantization of anti-entire functions also satisfies the property 
\begin{eqnarray*}
\widehat{\overline{z}}\psi \left( z\right) &=&T_{\hbar }\left( \overline{z}%
\right) \psi \left( z\right) =\int_{\mathbb{C}}e^{z\overline{w}/\hbar }%
\overline{w}\psi \left( w\right) d\mu _{\hbar }\left( w\right) \\
&=&\hbar \frac{d}{dz}\int_{\mathbb{C}}e^{z\overline{w}/\hbar }\psi \left(
w\right) d\mu _{\hbar }\left( w\right) \\
&=&\hbar \frac{d\psi \left( z\right) }{dz}.
\end{eqnarray*}
With this, we see that acting on the ground state $\Omega =1$, the operators 
$\widehat{z}$ and $\widehat{\overline{z}}$ ladder up and down, respectively,
the basis state $\left\{ z^{n}\right\} _{n\in Z}$. We thus define the
operators 
\begin{eqnarray}
A^{\dagger } &=&\widehat{z},  \label{ladder} \\
A &=&\widehat{\overline{z}},  \nonumber
\end{eqnarray}
and notice 
\[
\left[ A,A^{\dagger }\right] =\hbar . 
\]
We can now see that Toeplitz quantization is canonical. Specifically, 
\[
\left[ \widehat{x},\widehat{p}\right] =i\left[ A,A^{\dagger }\right] =i\hbar
. 
\]
Finally, we note that Toeplitz quantization gives an anti-Wick prescription
for operator ordering. For $\phi \in \mathcal{H}^{2}(\mathbb{C},d\mu _{\hbar
})$, 
\begin{eqnarray*}
T_{\hbar }\left( z^{m}\overline{z}^{n}\right) \psi \left( z\right) &=&\left(
\hbar \frac{d}{dz}\right) ^{n}\int_{\mathbb{C}}e^{z\overline{w}/\hbar
}w^{m}\psi \left( w\right) d\mu _{\hbar }\left( w\right) \\
&=&A^{n}\left( A^{\dagger }\right) ^{m}\psi \left( z\right) ,
\end{eqnarray*}
where we have used the fact that $\exp \left( z\overline{w}/\hbar \right) $
is the reproducing kernel.

We next begin an explicit construction of the quantum torus in this
representation. Let us define the operators analogous to the function $\exp
\left( 2\pi ix\right) $ and $\exp \left( 2\pi ip\right) $ which generate the
classical algebra of observables on the torus. We let 
\begin{eqnarray}
U &=&U\left( -i\hbar \pi \sqrt{2}\right)  \label{quantum algebra} \\
V &=&V\left( \hbar \pi \sqrt{2}\right) .  \nonumber
\end{eqnarray}
Notice from eqn. \ref{projective rep} that they obey the commutation
relations $UV=e^{i\lambda }VU$ with $\lambda =4\pi ^{2}\hbar .$ We can see
directly that eqn. \ref{quantum algebra} is the quantization of the
generators of the classical algebra. Observe 
\begin{eqnarray*}
e^{2\pi i\widehat{x}}\psi \left( z\right) &=&e^{\sqrt{2}\pi i\left(
A+A^{\dagger }\right) }\psi \left( A^{\dagger }\right) e^{-\sqrt{2}\pi
i\left( A+A^{\dagger }\right) }e^{\sqrt{2}\pi i\left( A+A^{\dagger }\right)
}\cdot 1 \\
&=&\psi \left( A^{\dagger }+i\hbar \pi \sqrt{2}\right) e^{\sqrt{2}\pi
iA^{\dagger }-\pi ^{2}\hbar }\cdot 1 \\
&=&e^{-\pi ^{2}\hbar +\sqrt{2}\pi iz}\psi \left( z+i\hbar \pi \sqrt{2}\right)
\\
&=&U\psi \left( z\right) .
\end{eqnarray*}
Likewise, 
\[
e^{2\pi i\widehat{p}}\psi \left( z\right) =V\psi \left( z\right) . 
\]
We call $U$ and $V$ the generators of the quantum algebra $\frak{A}_{\hbar }$
on the torus. More precisely, we have the following definition.

\begin{definition}
The (von Neuman) algebra $\frak{A}_{\hbar }$ is the weak closure of the set
of bounded operators generated by $\left\{ U,V\right\} $.
\end{definition}

\subsection{\protect\smallskip Wigner Representation of the Quantum Torus}

What is this algebra? In fact, the algebra is different for different values
of Planck's constant. Consider now the case of Planck's constant irrational.
(The rational case provides for another beautiful level of structure we
study in more detail later (for $h=1/N$).) For instance, does there exist a
bounded operator on $L^{2}\left( \mathbb{R}\right) $ which is not in $\frak{A%
}_{\hbar }$? The answer to this is yes, as can be seen from the fact that 
\[
\left[ U,e^{i\widehat{x}/\hbar }\right] =\left[ V,e^{i\widehat{x}/\hbar
}\right] =\left[ U,e^{i\widehat{p}/\hbar }\right] =\left[ V,e^{i\widehat{p}%
/\hbar }\right] =0 
\]
In other words, the algebra is contained in the commutant of the set of
bounded operators generated by $\exp \left( i\widehat{x}/\hbar \right) $and $%
\exp \left( i\widehat{p}/\hbar \right) $: $\frak{A}_{\hbar }\subset \left\{
e^{i\widehat{x}/\hbar },e^{i\widehat{p}/\hbar }\right\} ^{\prime }\equiv 
\frak{A}$. In fact, we can show that $\frak{A}_{\hbar }=\frak{A}$, and in
the process learn quite a bit about the properties of this algebra. We
follow here a suggestion of Coleman (ref. \cite{C}) to look at the Wigner
representation. It should be noted that there is a vast mathematical
literature and that the following calculation is by no means original. We
present here an argument for physicists which can be thought of as densely
filling the actual proof. For a full and rigorous presentation, the reader
is referred to ref. \cite{Ri}.

A von Neumann algebra $\frak{a}$ is a set of bounded operators equal to its
bicommutant: $\frak{a}=\frak{a}^{\prime \prime }$. A beautiful theorem of
von Neumann is that such an algebra is weakly closed. In our case, $\frak{A}=%
\frak{A}^{\prime \prime }$, since it is always true that $\frak{a}^{\prime }=%
\frak{a}^{\prime \prime \prime }$(see \cite{VN}). Thus by construction, $%
\frak{A}$ is a von Neumann algebra.

Recall the Wigner representation $a\left( p,q\right) $ of an operator $W$ is
given by 
\[
a\left( p,q\right) =\frac{1}{2\pi \hbar }\int e^{-ipx/\hbar }\left\langle
q+x/2\right| A\left| q-x/2\right\rangle dx, 
\]
for a $\psi \in L^{2}\left( \mathbb{R}\right) $. The Wigner representation
of the operator $W$ has the property that 
\[
\int a\left( p,q\right) dp=\left\langle q\right| A\left| q\right\rangle , 
\]
and 
\[
\int a\left( p,q\right) dq=\left\langle p\right| A\left| p\right\rangle . 
\]
Now, an operator in the commutant $\left\{ e^{i\widehat{x}/\hbar },e^{i%
\widehat{p}/\hbar }\right\} ^{\prime }$ satisfies 
\begin{eqnarray*}
e^{i\widehat{x}/\hbar }Ae^{-i\widehat{x}/\hbar } &=&A \\
e^{i\widehat{p}/\hbar }Ae^{-i\widehat{p}/\hbar } &=&A.
\end{eqnarray*}
What does this give for the Wigner functions? With a simple calculation, we
see 
\[
a\left( p,q\right) =a\left( p+1,q\right) =a\left( p,q+1\right) . 
\]
That is, the Wigner representation of an operator $A\in \frak{A}$ must be
periodic with period one.

For $h$ irrational, the weak closure of the set generated by $\left\{
U,V,e^{i\widehat{x}/\hbar },e^{i\widehat{p}/\hbar }\right\} $ comprise all
bounded operators on the Hilbert space $\frak{B}\left( \mathcal{H}\right) $.
To see this, we must simply show that the commutant of this set consists
only of the identity operator. (Note that $\frak{B}\left( \mathcal{H}\right)
=\frak{B}\left( \mathcal{H}\right) ^{\prime \prime }=\left\{ I\right\}
^{\prime }.$). We follow here an argument due to Faddeev \cite{Fa}.

We already know what the conditions are for an operator to commute with $%
\exp \left( i\widehat{x}/\hbar \right) $and $\exp \left( i\widehat{p}/\hbar
\right) $, that is, its Wigner functions must be periodic with period one.
In exactly the same we, we see that for an operator to commute with $U$ and $%
V$, we must have 
\[
a\left( p+h,q\right) =a\left( p,q+h\right) =a\left( p,q\right) . 
\]
Now the only function which is periodic with two periods irrationally
related is the constant function: $a\left( p,q\right) =a_{0}$, corresponding
to the operator $A=a_{0}I$.

We thus see that any operator in $A\in $ $\frak{B}\left( \mathcal{H}\right) $
can be written as the weak limit of finite polynomials of the form 
\begin{equation}
A_{i}=\sum_{a,b,c,d}\alpha _{a,b,c,d}U^{a}V^{b}\left( e^{i\widehat{x}/\hbar
}\right) ^{c}\left( e^{i\widehat{p}/\hbar }\right) ^{d}  \label{A}
\end{equation}
where the sum is taken to be finite. Now suppose further that $A_{i}\in 
\frak{A}$. Then for all $M$, 
\[
A_{i}=\frac{1}{M}\sum_{m=0}^{M-1}\left( e^{i\widehat{x}/\hbar }\right)
^{m}A_{i}\left( e^{i\widehat{x}/\hbar }\right) ^{-m}. 
\]
Substituting this into \ref{A}, we see 
\[
\sum_{a,b,c,d}\alpha _{a,b,c,d}U^{a}V^{b}\left( e^{i\widehat{x}/\hbar
}\right) ^{c}\left( e^{i\widehat{p}/\hbar }\right) ^{d}\left( 1-\frac{1}{M}%
\sum_{m=0}^{M-1}e^{2\pi imd/h}\right) =0\text{.} 
\]
As $M\rightarrow \infty $, the only way this can hold for $h$ irrational is
if $d=0$. In a similar fashion we see $c=0$. Thus for $A\in \frak{A}$, $A$
must be the weak limit of a finite polynomial operator of the form 
\[
A_{i}=\sum_{a,b}\alpha _{a,b}U^{a}V^{b} 
\]
which is precisely the von Neumann algebra $\frak{A}_{\hbar }$ defined
above. Thus we have demonstrated $\frak{A}_{\hbar }=\frak{A}.$

\subsection{Quantum Ergodicity on the Torus}

With this in mind, we can now give precisely what we mean by ``quantum
mixing'' and ``quantum ergodicity'' on the torus. We can define a ``quantum
phase average'' as a trace over the algebra $\frak{A}_{\hbar }$. Let $\phi
\in \mathcal{H}^{2}\left( \mathbb{C},d\mu _{\hbar }\right) $ be an arbitrary
vector of norm one. For $A\in \frak{A}_{\hbar }$, we define 
\begin{equation}
\tau _{\hbar }\left( A\right) =\int_{T^{2}}\left( U\left( l\right) \phi
,AU\left( l\right) \phi \right) d^{2}l\text{ .}  \label{trace}
\end{equation}
As shown in ref. \cite{KLMR}, this functional has the desired property that $%
\tau _{\hbar }\left( U^{m}V^{n}\right) =\delta _{m0}\delta _{n0}$. Another
fact about $\tau _{\hbar }$ is that its value on a Toeplitz operator is
equal to the integral of the symbol of the operator: 
\begin{equation}
\tau _{\hbar }\left( T_{\hbar }\left( f\right) \right) =\int_{T^{2}}f\left(
x,p\right) dxdp  \label{classical quantum}
\end{equation}
where $\tau \left( f\right) $ is the phase-space integral of $f$ over the
torus.

\begin{definition}
\label{quantum toral ergodicity}We define quantum ergodicity of a quantum
toral map $\left( \frak{A}_{\hbar }\text{, }\alpha \text{, }\mathbb{Z}\text{%
, }\tau _{\hbar }\right) $ to be the property 
\begin{equation}
s-\lim_{M\rightarrow \infty }\left\langle A\right\rangle _{M}=\tau _{\hbar
}\left( A\right)  \label{quantum ergodic}
\end{equation}
for any $A\in \frak{A}_{\hbar }$, where 
\[
\left\langle A\right\rangle _{M}=\frac{1}{M}\sum_{m=0}^{M-1}F^{m}AF^{-m}%
\text{.} 
\]
\end{definition}

\begin{definition}
\label{quantum toral mixing} Likewise, we define a quantum toral map to be
``quantum mixing'', if for any $A,B\in \frak{A}_{\hbar }$, 
\begin{equation}
\lim_{M\rightarrow \infty }\tau _{\hbar }\left( F^{M}AF^{-M}B\right) =\tau
_{\hbar }\left( A\right) \tau _{\hbar }\left( B\right) \text{.}
\label{quantum mixing}
\end{equation}
\end{definition}

\section{The $\theta $-torus}

It was shown in ref. \cite{KLMR} that a remarkable set of properties can be
associated with quantum dynamics on a torus if we let Planck's constant
satisfy the integrality condition 
\[
h=1/N. 
\]
We review these results briefly here and then present further work. The
algebra $\frak{A}_{\hbar }$ has a natural (and non-trivial) center generated
by 
\begin{eqnarray}
X &=&U^{N},  \label{center} \\
Y &=&V^{N}.  \nonumber
\end{eqnarray}
That is, for $h=1/N$, we easily see that 
\[
\left[ X,Y\right] =\left[ X,U\right] =\left[ X,V\right] =\left[ Y,U\right]
=\left[ Y,V\right] =0. 
\]
This insight was used to construct an $N$-dimensional ``quantum torus'' in
the following manner. The simultaneous eigenvalue problem: 
\begin{eqnarray*}
X\phi \left( z\right) &=&e^{2\pi i\theta _{1}}\phi \left( z\right) , \\
Y\phi \left( z\right) &=&e^{2\pi i\theta _{2}}\phi \left( z\right) ,
\end{eqnarray*}
where $\theta =\left( \theta _{1},\theta _{2}\right) \in \Bbb{T}^{2}$, was
used to decompose Bargmann space into the direct sum space 
\[
\kappa :\mathcal{H}^{2}\left( \mathbb{C},d\mu _{\hbar }\right) \rightarrow
\int_{T^{2}}^{\oplus }\mathcal{H}_{\hbar }\left( \theta \right) d\theta 
\]
where an element $\phi _{n}^{\left( \theta \right) }$ is a simultaneous
generalized eigenvectors of $X$ and $Y$, and $\mathcal{H}_{\hbar }\left(
\theta \right) $ denote the $N$-dimensional space of simultaneous
eigenvectors with fixed $\theta $. The isomorphism in fact preserves the
inner product, given by 
\begin{equation}
\left( \phi _{1},\phi _{2}\right) _{\mathcal{H}^{2}\left( \mathbb{C},d\mu
_{\hbar }\right) }=\int_{T^{2}}\int_{D}\overline{\kappa \phi _{1}\left(
z,\theta \right) }\kappa \phi _{2}\left( z,\theta \right) d\mu _{\hbar
}\left( z\right) d\theta \text{.}  \label{inner product}
\end{equation}
The space $\mathcal{H}_{\hbar }\left( \theta \right) $ thus has an inner
product defined as an integral over the fundamental domain $D=\left[
0,1\right] \times \left[ 0,1\right] $. A natural set of orthonormal position
state basis vectors exist for this given by the following functions: 
\[
\phi _{m}^{\left( \theta \right) }\left( z\right) =C_{m}\left( \theta
\right) e^{-N\pi z^{2}+2\sqrt{2}\pi \left( \theta _{1}+m\right) z}\vartheta
\left( -i\sqrt{2}Nz+i\left( \theta _{1}+i\theta _{2}+m\right) ,iN\right) 
\]
where 
\[
C_{m}\left( \theta \right) :=\left( 2/N\right) ^{1/4}e^{-\pi \left( \theta
_{1}+m\right) ^{2}/N-2\pi i\theta _{2}m/N}, 
\]
and 
\[
\vartheta \left( \omega ,\tau \right) =\sum_{k\in \mathbb{Z}}e^{i\pi
k^{2}\tau +2\pi ik\omega }. 
\]
is a Jacobi $\vartheta $-function (see, for example, \cite{Mu}). These
functions satisfy 
\[
\int_{D}\overline{\phi _{m}^{\left( \theta \right) }\left( z\right) }\phi
_{n}^{\left( \theta \right) }\left( z\right) d\mu _{\hbar }\left( z\right)
=\delta _{mn}\text{.} 
\]
We can see now why these functions are called ``position-state
eigenvectors''. Under the isomorphism $\kappa $, $U$ and $V$ are \textit{%
block diagonal}\textbf{, }that is, they leave a vector at a particular value
of $\theta $ at the same value, and furthermore, these $\phi _{m}^{\left(
\theta \right) }$ are also eigenvalues of $U^{n}$ for $n\in \left[
0,N-1\right] $: 
\begin{eqnarray}
\kappa U\kappa ^{-1}\phi _{m}\left( \theta ,z\right) &=&e^{2\pi i\left(
\theta _{1}+m\right) /N}\phi _{m}\left( \theta ,z\right) ,  \label{uandv} \\
\kappa V\kappa ^{-1}\phi _{m}\left( \theta ,z\right) &=&e^{2\pi i\theta
_{2}/N}\phi _{m+1}\left( \theta ,z\right) \,.  \nonumber
\end{eqnarray}

We are now in a position to calculate explicitly the form for the Hilbert
space of operators $L^{2}\left( \frak{A}_{\hbar },\tau _{\hbar }\right) $
for the quantized torus discussed in the introduction. We have the following
result.

\begin{theorem}
For the quantized torus, $L^{2}\left( \frak{A}_{\hbar },\tau _{\hbar
}\right) \simeq \frak{M}_{N\times N}\otimes L^{2}\left( \mathbb{T}%
^{2}\right) $.
\end{theorem}

Proof. Recall that any $A\in \frak{A}_{\hbar }$ can be written as the weak
closure of the finite sums 
\begin{equation}
\sum_{n,m}a_{n,m}U^{n}V^{m}=\sum_{c,d}\beta
_{c,d}X^{c}Y^{d}\sum_{a,b=0}^{N-1}\alpha _{a,b}U^{a}V^{b},
\label{Fourier series}
\end{equation}
where we have let $n=a+Nc$, $m=b+Nd$. Observe also that

\begin{eqnarray*}
\tau _{\hbar }\left( A^{\dagger }A\right) &=&\int_{\mathbb{T}^{2}}\left(
U\left( l\right) \phi ,A^{\dagger }AU\left( l\right) \phi \right) _{\mathcal{%
H}^{2}}d^{2}l \\
&=&\int_{\mathbb{T}^{2}}\left( U\left( l\right) \phi \right. , \\
&&\left. \left( \sum_{g,h\in \mathbb{Z}}\beta
_{g,h}^{*}X^{-g}Y^{-h}\sum_{e,f=0}^{N-1}\alpha _{e,f}^{*}U^{-e}V^{-f}\right)
\sum_{c,d\in \mathbb{Z}}\beta _{c,d}X^{c}Y^{d}\sum_{a,b=0}^{N-1}\alpha
_{a,b}U^{a}V^{b}U\left( l\right) \phi \right) _{\mathcal{H}^{2}}d^{2}l \\
&=&\int_{\mathbb{T}^{2}}\left( U\left( l\right) \phi ,\sum_{c,d,g,h\in %
\mathbb{Z}}\beta _{c,d}\beta
_{g,h}^{*}X^{c-g}Y^{d-h}\sum_{a,b,e,f=0}^{N-1}\alpha _{a,b}\alpha
_{e,f}^{*}U^{a-e}V^{b-f}U\left( l\right) \phi \right) _{\mathcal{H}%
^{2}}d^{2}l.
\end{eqnarray*}
We can write this as 
\begin{eqnarray*}
\tau _{\hbar }\left( A^{\dagger }A\right) &=&\int_{\mathbb{T}^{2}}\int_{%
\mathbb{T}^{2}}\left( \kappa U\left( l\right) \phi \left( \theta \right)
,\right. \\
&&\left. \kappa \sum_{c,d,g,h\in \mathbb{Z}}\beta _{c,d}\beta
_{g,h}^{*}X^{c-g}Y^{d-h}\sum_{a,b,e,f=0}^{N-1}\alpha _{a,b}\alpha
_{e,f}^{*}U^{a-e}V^{b-f}\kappa ^{-1}\kappa U\left( l\right) \phi \left(
\theta \right) \right) _{P}d^{2}\theta d^{2}l \\
&=&\int_{\mathbb{T}^{2}}\int_{\mathbb{T}^{2}}\left( \kappa U\left( l\right)
\phi \left( \theta \right) ,\kappa \sum_{a,b,e,f=0}^{N-1}\alpha _{a,b}\alpha
_{e,f}^{*}U^{a-e}V^{b-f}\kappa ^{-1}\kappa U\left( l\right) \phi \left(
\theta \right) \right) _{P}d^{2}l \\
&&\times \sum_{c,d,g,h\in \mathbb{Z}}\beta _{c,d}\beta _{g,h}^{*}e^{2\pi
i\left( c-g\right) \theta _{1}}e^{2\pi i\left( d-h\right) \theta
_{2}}d^{2}\theta \\
&=&\int_{\mathbb{T}^{2}}\left( U\left( l\right) \phi
,\sum_{a,b,e,f=0}^{N-1}\alpha _{a,b}\alpha _{e,f}^{*}U^{a-e}V^{b-f}U\left(
l\right) \phi \right) _{\mathcal{H}^{2}}d^{2}l \\
&&\times \int_{\mathbb{T}^{2}}\sum_{c,d,g,h\in \mathbb{Z}}\beta _{c,d}\beta
_{g,h}^{*}e^{2\pi i\left( c-g\right) \theta _{1}}e^{2\pi i\left( d-h\right)
\theta _{2}}d^{2}\theta .
\end{eqnarray*}
The set of operators $\left\{ U^{j}V^{k}\right\} $ in the last line of the
above expression form an $N^{2}$-dimensional vector space isomorphic to $%
\frak{M}_{N\times N}.$ From the second set of integrals it is clear that the
function $f\left( \theta \right) =$ $\sum_{c,d\in \mathbb{Z}}\beta
_{c,d}e^{2\pi ic\theta _{1}}e^{2\pi id\theta _{2}}$ span a dense subpace of $%
L^{2}\left( \mathbb{T}^{2}\right) $. The usual continuity argument completes
the proof of the theorem. 
\endproof%

\subsection{The Quantum Torus from $L^{2}\left( \mathbb{R}\right) $}

What we have demonstrated is a transformation of Bargmann space into a ``$%
\theta $-torus'' space. As there is an explicit isomorphism between $%
\mathcal{H}^{2}\left( \mathbb{C},d\mu _{\hbar }\right) $ and $L^{2}\left( %
\mathbb{R},dx\right) $, we can similarly construct the transformation
between $L^{2}\left( \mathbb{R},dx\right) $ and $\int_{T^{2}}^{\oplus }%
\mathcal{H}_{\hbar }\left( \theta \right) d\theta $. Applying the Bargmann
transformation to the basis functions $\phi _{m}^{(\theta )}\in \mathcal{H}%
(\theta )$, we find 
\[
\Phi _{m}^{(\theta )}(x)=B^{-1}\phi _{m}^{(\theta )}(x)=\frac{e^{2\pi
i\theta _{2}m/N}}{N^{1/2}}\sum_{k\in \mathbb{Z}}e^{2\pi i\theta _{2}k}\delta
\left( x-\frac{m+\theta _{1}+NK}{N}\right) . 
\]
We have thus found explicitly the $\delta $-comb wavefunctions described
informally in the physics literature (see, for example, \cite{HB}). For
later convenience, we express this now in Dirac notation as follows: 
\begin{equation}
\Phi _{m}^{(\theta )}=\frac{e^{2\pi i\theta _{2}m/N}}{N^{1/2}}\sum_{k\in %
\mathbb{Z}}e^{2\pi i\theta _{2}k}\left| \frac{m+\theta _{1}}{N}%
+k\right\rangle _{x}.  \label{x-rep}
\end{equation}

We can, of course, just as easily work in momentum representation. In fact,
for $h=1/N$, a rather interesting calculational identity can be found.

\begin{lemma}
For $h=1/N$, 
\[
\Phi _{m}^{(\theta )}=e^{-2\pi i\theta _{1}\theta _{2}/N}\sum_{n=0}^{N-1}%
\mathcal{F}_{mn}\widetilde{\Phi }_{n}^{(\theta )}, 
\]
where $\left\{ \widetilde{\Phi }_{n}^{(\theta )}\right\} _{0\leq n\leq N-1}$
are the momentum-state wave functions on the torus, 
\begin{equation}
\widetilde{\Phi }_{n}^{(\theta )}=\frac{e^{-2\pi in\theta _{1}/N}}{\sqrt{N}}%
\sum_{k}e^{-2\pi i\theta _{1}k}\left| \frac{\theta _{2}+n}{N}+k\right\rangle
_{p},
\end{equation}
and $\mathcal{F}_{mn}$ is the discrete Fourier transform, 
\[
\mathcal{F}_{mn}=\frac{e^{-2\pi imn/N}}{\sqrt{N}}. 
\]
\end{lemma}

\begin{remark}
We see in particular that for the subsets $\theta _{1}=0$ or $\theta _{2}=0$%
, changing coordinates from momentum representation to position
representation is simply a discrete Fourier transform.
\end{remark}

\proof%
The proof is a direct calculation. We have 
\begin{eqnarray*}
\int_{\mathbb{R}}\left| p\right\rangle \left\langle p\right| \Phi
_{m}^{(\theta )}dp &=&\frac{e^{2\pi i\theta _{2}m/N}}{N^{1/2}}\sum_{k\in %
\mathbb{Z}}e^{2\pi i\theta _{2}k}\int_{\mathbb{R}}\left| p\right\rangle
\left\langle p{\Huge |}\frac{m+\theta _{1}}{N}+k\right\rangle _{x}dp \\
&=&e^{2\pi i\theta _{2}m/N}\sum_{k\in \mathbb{Z}}e^{2\pi i\theta _{2}k}\int_{%
\mathbb{R}}\left| p\right\rangle \exp \left\{ -2\pi Nip\left( \frac{m+\theta
_{1}}{N}+k\right) \right\} dp \\
&=&e^{2\pi i\theta _{2}m/N}\int_{\mathbb{R}}\left| p\right\rangle e^{-2\pi
ip(m+\theta _{1})}\sum_{k\in \mathbb{Z}}e^{2\pi ik(\theta _{2}-Np)}dp \\
&=&\frac{e^{2\pi i\theta _{2}m/N}}{N}\int_{\mathbb{R}}\left| p\right\rangle
e^{-2\pi ip(m+\theta _{1})}\sum_{k\in \mathbb{Z}}\delta \left( p-\frac{%
\theta _{2}+k}{N}\right) dp \\
&=&\frac{e^{2\pi i\theta _{2}m/N}}{N}\sum_{k\in \mathbb{Z}}\left| \frac{%
\theta _{2}+k}{N}\right\rangle _{p}\exp \left\{ -2\pi i\left( \frac{\theta
_{2}+k}{N}\right) (m+\theta _{1})\right\} .
\end{eqnarray*}
We now let $k\rightarrow n+kN$, with $n\in \left\{ 0,...,N-1\right\} $, and $%
k\in \mathbb{Z}$, to find 
\begin{eqnarray*}
\int_{\mathbb{R}}\left| p\right\rangle \left\langle p\right| \Phi
_{m}^{(\theta )}dp &=&\frac{e^{2\pi i\theta _{2}m/N}}{N}\sum_{n}\sum_{k}%
\left| \frac{\theta _{2}+n}{N}+k\right\rangle _{p} \\
&&\times \exp \left\{ -2\pi i\left( \frac{\theta _{2}+n+Nk}{N}\right)
(m+\theta _{1})\right\} \\
&=&\frac{e^{-2\pi i\theta _{1}\theta _{2}/N}}{N}\sum_{n}e^{-2\pi
imn/N}e^{-2\pi in\theta _{1}/N}\sum_{k}e^{-2\pi i\theta _{1}k}\left| \frac{%
\theta _{2}+n}{N}+k\right\rangle _{p} \\
&=&e^{-2\pi i\theta _{1}\theta _{2}/N}\sum_{n}\mathcal{F}_{mn}\widetilde{%
\Phi }_{n}^{(\theta )},
\end{eqnarray*}
as claimed. 
\endproof%

Analogous to eqn. \ref{inner product}, we can also find an explicit
expression for the inner product over the $N$-dimensional Hilbert space at
each point on the $\theta $-torus as an integral over the fundamental domain 
$[0,1]$ of the real line. The inner product defined in eqn. \ref{inner
product} can be written as 
\[
\left( \Psi _{1}(\theta ),\Psi _{2}(\theta )\right) _{P}=\int_{0}^{1}%
\overline{\Psi _{1}(x,\theta )}(K\Psi _{2})(x,\theta )dx, 
\]
where 
\[
K(x,y)=g\left( \frac{x-y}{2\hbar }\right) , 
\]
and 
\[
g(r)=\frac{1}{2\pi \hbar }e^{-\hbar r^{2}+ir}\frac{\sin r}{r} 
\]

\proof%
The proof is again a direct calculation. We have 
\begin{eqnarray*}
\int_{D}\overline{\psi _{1}(z,\theta )}\psi _{2}(z,\theta )d\mu _{\hbar }(z)
&=&\int_{D}\overline{B\Psi _{1}(z,\theta )}\Psi _{2}(z,\theta )d\mu _{\hbar
}(z) \\
&=&\frac{1}{2\left( \pi \hbar \right) ^{3/2}}\int_{\mathbb{R}^{2}}\overline{%
\Psi _{1}(x,\theta )}\Psi _{2}(y,\theta )e^{-(x^{2}+y^{2})/2\hbar } \\
&&\times \int_{D}e^{-(u^{2}-x(u-iv)+y(u+iv))/\hbar }dudvdxdy \\
&=&\frac{1}{2\left( \pi \hbar \right) ^{3/2}}\sum_{k\in \mathbb{Z}%
}\int_{0}^{1}dx\int_{-\infty }^{\infty }dy\overline{\Psi _{1}(x+k,\theta )}%
\Psi _{2}(y,\theta )e^{-((x+k)^{2}+y^{2})/2\hbar } \\
&&\times \int_{D}e^{-(u^{2}-(x+k)(u-iv)+y(u+iv))/\hbar }dudv.
\end{eqnarray*}
We next use the fact that both $\Psi _{1}$ and $\Psi _{2}$ satisfy $X\Psi
_{i}=e^{2\pi i\theta _{1}}\Psi _{i}$ and $Y\Psi _{i}=e^{2\pi i\theta
_{2}}\Psi _{i}$. Substituting in, we find 
\begin{eqnarray*}
&&\frac{1}{2\left( \pi \hbar \right) ^{3/2}}\sum_{k\in \mathbb{Z}%
}\int_{0}^{1}dx\int_{-\infty }^{\infty }dye^{-2\pi ik\theta _{2}}\overline{%
\Psi _{1}(x,\theta )}\Psi _{2}(y,\theta )e^{-((x+k)^{2}+y^{2})/2\hbar } \\
&&\times \int_{D}dudve^{-(u^{2}-(x+k)(u-iv)+y(u+iv))/\hbar } \\
&=&\frac{1}{2\left( \pi \hbar \right) ^{3/2}}\sum_{k\in \mathbb{Z}%
}\int_{0}^{1}dx\int_{-\infty }^{\infty }dy\overline{\Psi _{1}(x,\theta )}%
\Psi _{2}(y-k,\theta )e^{-((x+k)^{2}+y^{2})/2\hbar } \\
&&\times \int_{D}e^{-(u^{2}-(x+k)(u-iv)+y(u+iv))/\hbar }dudv \\
&=&\frac{1}{2\left( \pi \hbar \right) ^{3/2}}\int_{0}^{1}dx\overline{\Psi
_{1}(x,\theta )}\int_{-\infty }^{\infty }dy\Psi _{2}(y,\theta
)e^{-(x^{2}+y^{2}+i(x-y))/2\hbar }\frac{\sin \left( \frac{x-y}{2\hbar }%
\right) }{\left( \frac{x-y}{2\hbar }\right) } \\
&&\times \int_{-\infty }^{\infty }e^{-((u-k)^{2}-(x+y)(u-k))/\hbar }du \\
&=&\int_{0}^{1}\overline{\Psi _{1}(x,\theta )}(K\Psi _{2})(x,\theta )dx,
\end{eqnarray*}
and the claim follows. 
\endproof%

We see that the kernel $K(x,y)$ is a type of quantum diffraction in keeping
with the uncertainty principle. In fact it can be readily seen that as $%
\hbar \rightarrow 0$, $K(x,y)\rightarrow \delta (x-y)$. We demonstrate this
numerically for $h=1/10$ and $h=1/100$. 
\[
.\FRAME{itbpFU}{3in}{2.0003in}{0in}{\Qcb{$\left| g\left( r/2\hbar \right)
\right| ^{2}$ for $h=1/10$.}}{\Qlb{diffraction}}{Diffraction }{%
\special{language "Scientific Word";type "MAPLEPLOT";width 3in;height
2.0003in;depth 0in;display "USEDEF";plot_snapshots TRUE;function
\TEXUX{$\left( \frac{1}{2\pi \left( 1/10\right) }\right) ^{2}e^{-2\left(
1/10\right) \left( r/\left( 2/10\right) \right) ^{2}}\left( \frac{\sin
r/\left( 2/10\right) }{r/\left( 2/10\right) }\right) ^{2}$};linecolor
"black";linestyle 1;linethickness 1;pointstyle "point";xmin "-5";xmax
"5";xviewmin "-5";xviewmax "5";yviewmin "-0.05066";yviewmax "2.585";phi
45;theta 45;plottype 4;numpoints 49;axesstyle "normal";xis
\TEXUX{r};var1name \TEXUX{$x$};valid_file "T";tempfilename
'C:/WINDOWS/TEMP/EWF6VF00.wmf';tempfile-properties "XP";}}\FRAME{itbpFU}{3in%
}{2.0003in}{0in}{\Qcb{$\left| g\left( r/2\hbar \right) \right| ^{2}$ for $%
h=1/100$.}}{\Qlb{diffraction2}}{diffraction2}{\special{language "Scientific
Word";type "MAPLEPLOT";width 3in;height 2.0003in;depth 0in;display
"USEDEF";plot_snapshots TRUE;function \TEXUX{$\left( \frac{1}{2\pi \left(
1/100\right) }\right) ^{2}e^{-2\left( 1/100\right) \left( r/\left(
2/100\right) \right) ^{2}}\left( \frac{\sin r/\left( 2/100\right) }{r/\left(
2/100\right) }\right) ^{2}$};linecolor "black";linestyle 1;linethickness
1;pointstyle "point";xmin "-5";xmax "5";xviewmin "-5";xviewmax "5";yviewmin
"-5.066";yviewmax "258.5";phi 45;theta 45;plottype 4;numpoints 49;axesstyle
"normal";xis \TEXUX{r};var1name \TEXUX{$x$};valid_file "T";tempfilename
'C:/WINDOWS/TEMP/EWF6VF01.wmf';tempfile-properties "XP";}} 
\]
Observe also that with respect to this inner product, the basis elements $%
\left\{ \Phi _{m}^{(\theta )}\right\} $ are orthonormal:

\begin{corollary}
$\left( \Phi _{m}^{(\theta )},\Phi _{n}^{(\theta )}\right) =\delta _{mn}.$
\end{corollary}

Proof. The proof is a straightforward calculation, but also follows
immediately from the Bargmann transformation on the inner product eqn. \ref
{inner product}. 
\endproof%

\end{document}